\documentclass[preprint2]{emulateapj}

\shorttitle{$S^2$ models of ETGs}
\shortauthors{Coppola et al.}

\begin{document}
\title{Sersic galaxy with Sersic halo models of early-type galaxies: a tool for N-body simulations}

\author{Coppola, G.,\altaffilmark{1},
        La Barbera, F.,\altaffilmark{2},
        Capaccioli, M.\altaffilmark{1,3}}
\affil{$^{(1)}$ Dip. di Scienze Fisiche, University of Naples Federico II (Italy)
}
\affil{$^{(2)}$INAF-Osservatorio Astronomico di Capodimonte, Via Moiariello 16, 80131 Napoli, ITALY
}
\affil{$^{(3)}$$INAF-VSTceN$, via Moiariello 16, 80131 Napoli, ITALY
}

\begin{abstract}
We present spherical, non-rotating, isotropic models of early-type galaxies with stellar and dark-matter components both described by deprojected Sersic density profiles, and prove that they represent physically admissible stable systems. Using empirical correlations and recent results of N-body simulations, all the free parameters of the models are expressed as functions of one single quantity: the total (B-band) luminosity of the stellar component.

We analyze how to perform discrete N-body realizations of Sersic models. To this end, an optimal smoothing length is derived, defined as the softening parameter minimizing the error on the gravitational potential for the deprojected Sersic model. It is shown to depend on the Sersic index $n$ and on the number of particles of the N-body realization.

A software code allowing the computations of the relevant quantities
of one- and two-component Sersic models is provided.  Both the code and the results of the present work are primarily intended as tools to perform N-body simulations of early-type galaxies, where the structural non-homology of these systems (i.e. the variation of the shape parameter along the galaxy sequence) might be taken into account.
\end{abstract}


\section{Introduction}~\label{intro}
Merging of red-sequence galaxies might be  an  important channel  for the
formation  of  massive  early-type galaxies  (ETGs).  Such \textit{dry}
mergers  have been observed to take place and have an impact on the population of ETGs at both low 
(up to $z \sim 0.3$; \citealt{whitaker2008},~\citealt{masjedi2008}), and intermediate redshift, in  cluster and field environments~\citep{vandokkum1999,vandokkum2005,tran2005,bell2006}.
A further evidence comes from the fact that the stellar mass on the red sequence has 
been found to be nearly doubled from  $z \sim 1$~\citep{zucca2006, bell2004} on, implying
that at least some red galaxies must be formed from merging systems that
are  either  very  dusty  or  gas-poor~\citep{faber2005}.
K-band selected  samples also revealed  a substantial  population of  old,
passively evolving, massive ETGs  already in place at $1<z<2$, with  luminosity
and  stellar mass functions evolving only  weakly up to
$z \sim 0.8-1$~\citep{cimatti2002, bundy2006, cimatti2006}.\\
From the theoretical viewpoint, dry mergers are also expected
to play a major role.  Using semi-analytical  models,
\citet{khochfar2003}  found that a  large fraction of present-day ETGs are indeed
formed by merging bulge-dominated  systems and  that the fraction  of spheroidal
mergers increases  with luminosity, with massive ETGs  being formed by
nearly dissipationless events.  As shown by~\citet{delucia2006},  more massive
ETGs are expected to be built up of several stellar pieces, with the number of
effective stellar progenitors increasing up to five for the most massive
galaxies. On the other hand, hydro-dynamical simulations have also
shown that accretion of smaller disk-dominated galaxies (in the mass ratio of 1:10) could also have an important
role in the evolution of massive ETGs, explaining the presence of the tidal debris observed
at $z \sim0$~\citep{FMC08}.

To constrain the  role  of dry  mergers in  galaxy
formation,  it is  of  importance to  perform
merging  simulations of spheroidal systems,  comparing the properties of merger remnants to
observations.  So far, merging simulations of ETGs have been mostly used
to constrain the origin of the empirical correlations  among galaxy's observed quantities,
such as the  Faber-Jackson~\citep[hereafter     $FJ$]{faber&jackson1976ApJ},     the     Kormendy~\citep[hereafter  KR]{kormendy1977ApJ},   and  the  Fundamental  Plane~\citep[hereafter FP]{djorgovski&davies1987ApJ}  relations. The impact
of  dry merging  has  been  investigated in  several
works (e.g.~\citealt{capelatoetal1995ApJ,                  dantasetal2003MNRAS,
  evstigneevaetal2004MNRAS,  nipotietal2003MNRAS}).  They have all agreed  that  dissipationless  merging  is  able to  move galaxies along the  FP.  But it is not clear if dry mergers are also able to preserve other
observed  correlations~\citep{Boylan06}. For instance, ~\cite{nipotietal2003MNRAS}
found that the products of repeated  merging of gas-free  galaxies are
characterized  by  an unrealistically large effective radius and a mass-independent
velocity dispersion, while~\citet{evstigneevaetal2004MNRAS} found that
only the merging of massive galaxies, that lie on the KR, leads to end-products that still
follow that relation. \\\
In previous works, merging simulations have been performed by means of ETG's
models  where the stellar  component is  described by  simple analytic density
laws,  such as  the King  or  the Hernquist  profiles.  This  approach
implicitly  neglects  one key  observational  feature: the  structural
non-homology of the  ETG population~\citep{GrC97}.  It is well  established that the
observed light profiles of ETGs deviate from a pure $r^{1/4}$ law, being
better described         by         the Sersic (\citeyear{s1968adga.book})                         model~\citep{caonetal1993MNRAS,  donofrioetal1994MNRAS,  grahametal1996ApJ}.
The Sersic index (shape parameter), $n$, measuring the steepness
of  the  light  profile,   changes  systematically  along  the  galaxy
sequence, the more  luminous galaxies having higher $n$. Moreover,
the shape  parameter also correlates with  other observed
properties of ETGs,  such as the effective parameters  and the central
velocity dispersion~\citep{graham2002}, as expected in view of the correlation of $n$ with the luminosity. Different values of $n$ correspond
to physical systems that differ  significantly in their phase-space density structure, with
higher Sersic indices describing galaxies whose light profile is significantly
more concentrated toward the center, with an extended low surface brightness halo.
Thus, merging systems with different $n$'s might lead to a different evolution
of the phase-space density of merging remnants with respect to that of ``homologous'' King/Hernquist
models.  For
what concerns dark matter  haloes, previous simulations have usually adopted
either  the Navarro-Frenk-White (NFW)  profile~\citep{navarroetal1995} or  the
Hernquist (\citeyear{hernquist1990ApJ}) profile.  However, as shown  by ~\citet{merritt2005} and ~\citet{merritt2006} (hereafter MGM06),  galaxy- and cluster-sized  halos are
actually  better  described  by   using  either  the  Einasto's  model~\citep{einasto1968}
or the  Prugniel \& Simien model~\citep{prugniel1997}
rather than a NFW-like profile~\citep{navarroetal1995}. The Einasto's model is identical
in  functional form  to  the  Sersic  model, but  is used  to describe the
deprojected  (rather than the projected) density profile, while   the Prugniel
\&  Simien model  is an analytic  approximation  to the  deprojected
Sersic profile. ~\citet{merritt2005} (hereafter MNL05) and MGM06 found that the deprojected Sersic model (i.e. the Prugniel \&  Simien model) provides a better fit to the projected mass density profile of simulated dark-matter halos, with a Sersic index value of $n \sim 3$ for galaxy-sized dark-matter halos.

Hence, the deprojected Sersic model seems able to describe both the stellar and dark matter components of ETGs.
Driven by that, we  present here new simple  models of ETGs, where both  components follow
the deprojected Sersic law. Hereafter, we refer to these models as double Sersic ($S^2$) models. The
models  describe  spherical,  non-rotating, isotropic  systems, and are
intended as a tool to perform N-body simulations of ETGs. In a companion contribution (Coppola et al.~2009b, in preparation), we use the $S^2$ models to investigate how
dissipation-less (major and minor) mergers affect the structural properties of ETGs, such as the shape of their light profile and their stellar population gradients.
The present paper aims at:  (i) describing the main characteristics of the
$S^2$  models, by deriving the corresponding potential-density pair and
distribution function (Sec.~2), and discussing  their physical {\it consistency}  and stability (Sec.~3);
(ii) describing how to perform discrete  N-body realization of the models, by  adopting  an optimal gravitational  smoothing   length for simulation codes (Sec.~4); (iii) giving a set of recipes to
fix all the free model parameters (Sec.~5); (iv)  providing the  software code to compute dynamical/structural properties of both the one- and two-component Sersic models. Summary and discussion are drawn in Sec.~6.

\section{The double Sersic ($S^2$) model}\label{model}

\subsection{The deprojected Sersic model}\label{submodel.1}
The surface brightness profile of ETGs, $I(R)$, is accurately described
by the Sersic law~\citep{capaccioli1992, caonetal1993MNRAS,donofrioetal1994MNRAS}:
\begin{equation} \label{eq_sersic}
    I(R; n) =I_{0} \exp\left[ - b \,(R/R_{e_L})^{1/n} \right] \; ,
\end{equation}
where $I_{0}$ is the central  surface brightness, $R$   is   the
(equivalent) projected distance to the galaxy center, $n$ is the
Sersic index (shape parameter), and $b$ is a function of
$n$, defined in  such a way  that $R_{e_L}$ is the effective (half-light) radius of the galaxy~\citep{ciotti1991A&A,ciotti1999}. The quantity $b$ is approximated at better than 1$\%$ by the relation $b\sim exp \left[ 0.6950+\ln(n)-0.1789/n \right]$~\citep{LN99}.

For a spherical system, under the assumption that the stellar
mass-to-light ratio, ${M}_{_{L}}/L$, does not change with radius, the spatial mass density  profile of the  stellar component, $\rho_{_{L}}$,  is obtained by solving the  Abel integral equation  (\citealt{binney&tremaine}),
\begin{equation}
\rho_{_{L}}(r) = -\frac{1}{\pi}  \frac{ {M}_{_{L}}}{L} \int_{r}^{\infty} \frac{d I}{d  R}   \,
\frac{d R}{\sqrt{R^{2}-r^{2}}}
\end{equation}
where  $r$ is  the distance to the galaxy center. 
Setting $u=r/R$ and inserting Eq.~\ref{eq_sersic} into the Abel equation, one obtains the following expression:
\begin{eqnarray}
\rho_{_{L}}(r; n) & = & \rho_{0_{L}} \; \widetilde{\rho}(x; n) =  \nonumber  \\
            &  & \hspace{-1cm} \rho_{0_{L}} \frac{b}{\pi n} x^{\frac{1}{n}-1} \int_{0}^{1} \frac{u^{-1/n} \,
    \exp[-b x^{1/n} u^{-1/n} ] \; du} {\sqrt{1-u^{2}}},   \;
\label{eq_dens_definitiva}
\end{eqnarray}
where $x=r/R_{e_L}$ is the distance to the galaxy center in units of $R_{e_L}$, $\widetilde{\rho}(x; n)$ is the dimensionless deprojected density profile, and $\rho_{0_{L}}= {M}_{_{L}}/R_{e_L}^3 \cdot  b^{2 n}/ (2 \pi n \, \Gamma(2n))$ is the scaling factor of the stellar density profile. Here, $\Gamma$ denotes the complete gamma function, and the expression of $\rho_{0_{L}}$ is obtained by using eq.~4 of~\citet{ciotti1999}, which gives the total luminosity of the Sersic model as a function of $I_0$, $R_{e_L}$, and $n$. From Eq.~\ref{eq_dens_definitiva}, one obtains the mass profile:
\begin{eqnarray}\label{eq_mass_definitiva}
    {M}_{_{L}}(r; n) & = & {M}_{0_{L}}  \widetilde{M}(x; n) = \nonumber \\
 & \hspace{-2.2cm} M_{0_{L}} & \hspace{-1.0cm} \frac{4}{b^{2n}} \int_{0}^{1} \frac{u^{2}}    {(1-u^{2})^{1/2}} \; \gamma\left[2n+1, b \left(\frac{x}{u}\right)^{1/n}
    \right] du, \;
\end{eqnarray}
where $\widetilde{M}$ is the dimensionless mass profile, and ${M}_{0_{L}}={M}_{_{L}} b^{2 n}/(2 \pi n \, \Gamma(2n))$ is the scaling factor of ${M}_{_{L}}(r)$. 
From the Laplace equation, one finds the following expression for the gravitational potential:
\begin{eqnarray}   \label{eq_potential_definitiva}
\varphi_{_{L}}(r; n) & = & \varphi_{0_{L}} \, \widetilde{\varphi}(x; n) =  - \varphi_{0_{L}} \frac{{\widetilde{M}}(x; n)}{x}  + \nonumber \\
 & - & \varphi_{0_{L}} \frac{4}{b^{n}}\int_{0}^{1} u (1-u^{2})^{-\frac{1}{2}}  \gamma \left( n+1, b \left(\frac{x}{u} \right)^{\frac{1}{n}} \right) du, \;
\end{eqnarray}
where $\widetilde{\varphi}(x; n)$ is the dimensionless gravitational potential, and
$\varphi_{0_{L}}={G {M}_{_{L}}}/{R_{e_L}}$ is the corresponding scaling factor.
As shown in Secs.~2.2 and~2.3, the above equations provide the essential ingredients to construct the $S^2$ models.

We notice that, due to the existence of radial gradients in stellar population  properties (such as age and metallicity) of ETGs (e.g.~\citealt{peletieretal1990AJ}), the assumption of a constant mass-to-light ratio (${M}_{_{L}}/L(r)=const.$) might not actually reflect the physical properties of early-type systems. As discussed in Sec.~\ref{sec_2_7}, considering the observational results on age and metallicity gradients in ETGs, the ${M}_{_{L}}/L$ is expected to vary significantly with galaxy radius (up to $\sim 50\%$) at optical wavebands (B-band). However, the variation is significantly reduced, becoming consistent with zero within observational uncertainties, at Near-Infrared (NIR) wavebands. According to that, we implicitly assume here that the parameters $R_{e_L}$ and $n$, entering the normalization factors of the potential--density pair of the $S^2$ models,  are those describing the NIR profile of ETGs.  In Sec.~\ref{sec_2_3}, we describe how to derive the free parameters of the $S^2$ models according to this assumption.

The deprojection of the Sersic law has been already presented in several works (\citealt{ciotti1991A&A, prugniel1997, mazure2002, terzic2005}). Following ~\citet{MM87}, ~\citet{prugniel1997}  provided an analytical approximation to the spatial density profile of the $R^{1/n}$ model (Eq.~\ref{eq_dens_definitiva}). \citet{LN99} showed that the Prugniel \& Simien approximation reproduces the deprojected Sersic profile with an accuracy better than 5$\%$, in the radial range of  $10^{-2}$ to $10^{3} R_{e_L}$, for Sersic indices between $n \sim 0.5$ and $n \sim 10$. The Prugniel \& Simien model has been also adopted by \citet{terzic2005} to present one-component Sersic models of ETGs with  power-law cores. Exact solutions to the deprojection of the $R^{1/n}$ model have been provided by \citet{mazure2002}, in terms of the so-called Meijer G functions, while  \citet{ciotti1991A&A} presented exact numerical expressions for the mass, gravitational potential, and central velocity dispersion of the one-component Sersic model. In the present work, we report a concise reference to the integral equations that define the density-potential pair, the mass profile and the distribution function of the deprojected  Sersic law. All the quantities 
characterizing the Sersic model can be numerically computed by using a set of publicly available Fortran programs (see App.~A).

\subsection{The dark matter Sersic model}\label{submodel.2}
MNL05 and~MGM06 found that the deprojected Sersic law provides a better fit to the density profile of dark matter halos than the NFW law. MNL05 found that a  Sersic index value of  $n=3.00 \pm 0.17$ is required to fit the profile of galaxy-sized halos. On the other hand, MGM06 fitted the Prugniel \& Simien model to the density profiles of galaxy-sized halos, finding a best-fitting value  of   
$n \sim  3.59 \pm 0.65$. Considering the lower uncertainty of the MNL05 estimate,  we describe the dark matter component of the models with a deprojected Sersic  model having $n=3$. The  corresponding density-potential pair and mass profile are then obtained from the equations:
\begin{eqnarray}
\rho_{_{D}}(r) & = &\frac{\mu}{x_{_{D}}^3} \; \rho_{0_{L}} \; \widetilde{\rho}\left( \frac{x}{x_{_{D}}} ; n=3\right) \label{eqs_DM1}\\
{M}_{_{D}}(r) & = & \mu \; {M}_{0_{L}} \; \widetilde{M}\left( \frac{x}{x_{_{D}}}; n=3 \right) \label{eqs_DM2} \\
\varphi_{_{D}}(r) & = & \frac{\mu}{x_{_{D}}}  \; \varphi_{0_{L}} \; \widetilde{\varphi}\left( \frac{x}{x_{_{D}}} ; n=3 \right), \label{eqs_DM3}
\end{eqnarray}
where the dimensionless density-potential pair ($\widetilde{\rho}$, $\widetilde{\varphi}$) and the dimensionless mass profile $\widetilde{M}$ are obtained by setting $n=3$ in Eqs.~\ref{eq_dens_definitiva},~\ref{eq_mass_definitiva} and~\ref{eq_potential_definitiva}, respectively. Here, we have denoted as $\mu = {{M}_{_{D}}}/{{M}_{_{L}}}$ the ratio of the total halo mass, ${M}_{_{D}}$, to the total stellar mass ${M}_{_{L}}$,  and $x_{_{D}}={R_{e_{D}}}/{R_{e_{L}}}$ the ratio of the (projected) effective radii of the dark matter and stellar components.

We notice that although we fix here the shape parameter value of the dark-matter component, the $S^2$ models could be directly generalized to the case where the Sersic index of the halo component changes with its mass~\footnote{To this aim, one should change Eqs.~\ref{eqs_DM1},~\ref{eqs_DM2}, and \ref{eqs_DM3}, by replacing the value of $n=3$ with a different Sersic index of the dark matter halo, and derive the distribution function of the model accordingly Sec.~\ref{DF}.}. Such a dependece is somewhat suggested by the results of MNL05 and~MGM06, who found that cluster-sized halos ($M_{_D} \sim 10^{15} M_{\odot}$) are better described with Sersic index values of $2.38 \pm 0.25$ and $\sim 2.89 \pm 0.49$, respectively,  these values being systematically smaller than those obtained for galaxy-sized halos. However, one should notice that, when fitting dwarf-sized dark matter halos ($M_{_D} \sim 10^{10} M_{\odot}$), MNL05 found  a best-fitting Sersic index value of $3.11 \pm 0.05$, which is fully consistent with that of $3.00 \pm 0.17$ found for galaxy-sized halos ($M_{_D} \sim 10^{12} M_{\odot}$). Hence, current results seem to suggest a very similar Sersic index value of $\sim 3$ for galaxy-sized halos of different masses, supporting our assumption of a fixed $n$ value.

\subsection{Density-potential pair and Distribution Function}~\label{DF}

The total mass density profile is obtained by adding up the profiles of the stellar and dark matter components:
\begin{equation}
\rho(r) = \rho_{_{L}} \! + \! \rho_{_{D}}= \rho_{0_{L}} \left[ \; \widetilde{\rho}(x; n) +  \frac{\mu}{x_{_{D}}^3} \; \widetilde{\rho}\left( \frac{x}{x_{_{D}}} ; 3\right) \right].
\end{equation}
From the linearity of the Laplace equation, the total gravitational potential is equal to $\varphi(r) = \varphi_{_{L}} \! + \! \varphi_{_{D}}$, where $\varphi_{_{L}}$ and $\varphi_{_{D}}$ are obtained from Eqs.~\ref{eq_potential_definitiva} and~\ref{eqs_DM3}. A similar expression
can also be obtained for the mass profile, combining Eqs.~\ref{eq_mass_definitiva} and~\ref{eqs_DM2}.  We note that the global density-potential pair  and the mass profile are completely defined from five parameters, which are the dimensional quantities ${M}_{_{L}}$ and $R_{e_{L}}$, and the  dimension-less parameters $x_{_{D}}$, $\mu$, and $n$.

The distribution function of a stationary, spherical, isotropic system depends only on the binding energy $E$ and is uniquely defined by the density-potential  pair
through the Eddington formula~\footnote{As usually done, we write the Eddington formula by adopting natural units, where $M_{_L}=1$, $R_{e_L}=1$, and $G=1$, with $G$ being the gravitational constant.} (\citealt{binney&tremaine}):
\begin{equation}\label{eq_distribution_function}
    f(\mathcal{E}) = \frac{1}{\sqrt{8} \pi^{2}} \left[ \int_{0}^{\mathcal{E}}
    \frac{d^{2} \rho}{d \Psi^{2}} \frac{d \Psi}{\sqrt{\mathcal{E} -
    \Psi}} + \frac{1}{\sqrt{\mathcal{E}}} \left ( \frac{d \rho}{d \Psi}
  \right)_{\Psi=0} \right] \; ,
\end{equation}
where $\Psi(r)  \equiv - \varphi(r) + \varphi_0$ and $\mathcal{E} \equiv -E + \varphi_0$ is the relative binding energy, with $\varphi_0$ being a suitably defined constant~\citep[see][]{binney&tremaine}.
For the $S^2$ models, the global potential and density profiles are proportional to the dimensional factors $\varphi_{0_L}$ (see Eqs.~5 and~8) and $\rho_{0_L}$ (see Eqs.~3 and~6). Hence, using Eq.~B1 in
 App.~\ref{DE2SERSIC}, one finds that, unless of a scaling factor depending on $M_{_L}$ and $R_{e_L}$,  the $f(\mathcal{E})$ is determined by the three dimension-less parameters $x_{_{D}}$, $\mu$, and $n$. As for the case of single Sersic models~\citep{ciotti1991A&A}, one can show that the second term on the right side of Eq.~\ref{eq_distribution_function} is always equal to zero for all possible values of $x_{_{D}}$, $\mu$, and $n$. In fact, one can write $(d \rho /  d\Psi)_{\Psi=0} = \lim_{r \rightarrow \infty}  (d \rho / d r) (dr/d\Psi)$. For $r \rightarrow \infty$, the first derivative of the gravitational potential  decreases as $r^{-2}$, while the first derivative of the density decreases exponentially (see eq.~8 of \citealt{ciotti1991A&A}), implying that $\lim_{r \rightarrow \infty}  (d \rho / d r) (dr/d\Psi)=0$.
In App.~\ref{DE2SERSIC}, we report in detail how to calculate the distribution function by expressing the function $\frac{d^{2} \rho}{d \Psi^{2}}$ in terms of the first and second derivatives of $\widetilde{\rho}$, the gravitational potential $\widetilde{\varphi}$, and the mass profile of the dark matter and stellar components.

\section{Physical consistency and stability}\label{consistency}

The  Eddington inversion  does not  guarantee that the distribution function
is a physically admissible  stationary solution of the Boltzmann equation.
To this effect, for a given density-potential pair, one has to show
that $f(\mathcal{E})$  is non-negative for all positive values of the relative binding energy. As shown by~\cite{ciotti1991A&A}, one-component spherical, non-rotating, isotropic Sersic models are always physically admissible, while in the anisotropic case, a minimum
anisotropy radius exists for the model to be admissible, with this radius
depending on the Sersic index $n$~\citep{ciotti1997}.

\begin{figure*}[!t]
    \centering
        \includegraphics[width=12cm,height=16cm]{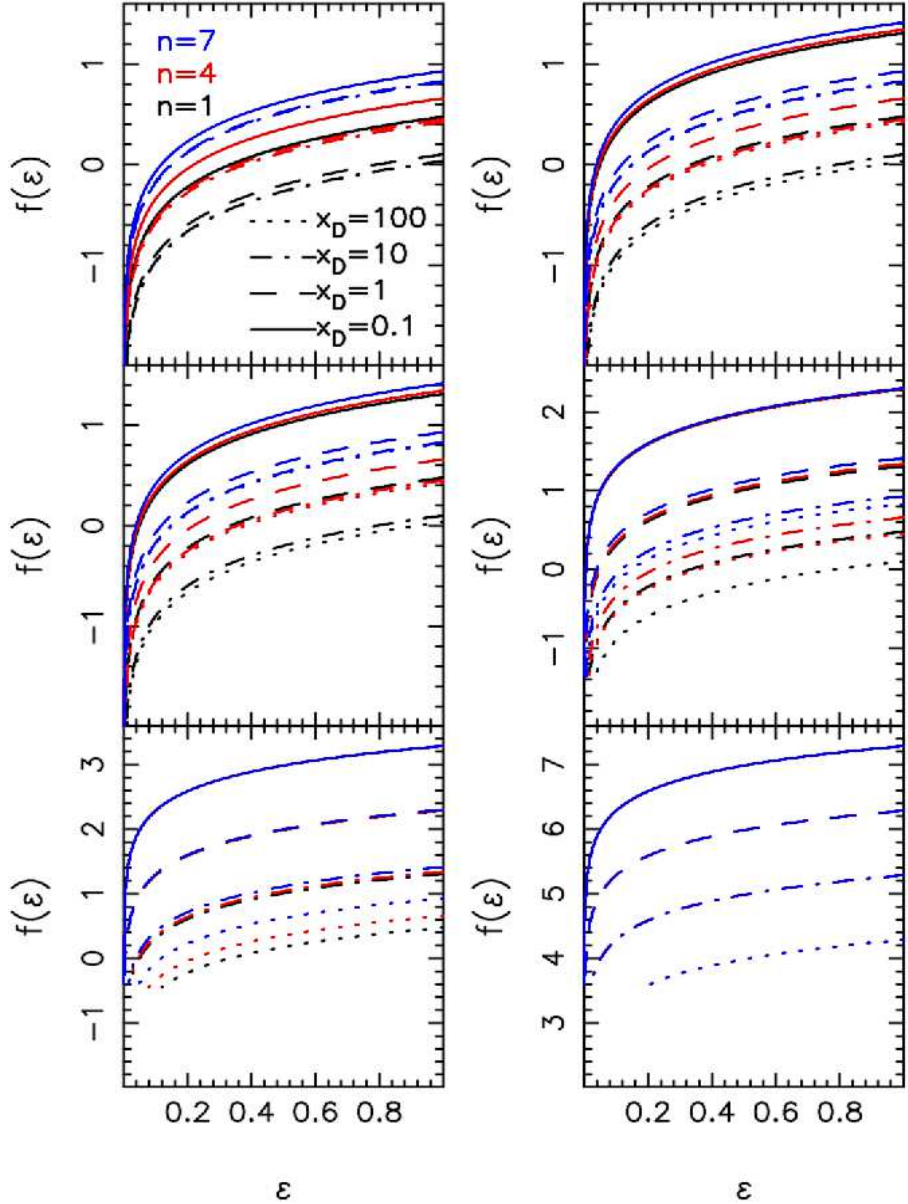}
                \caption{\footnotesize{Physical {\it consistency} of the $S^2$ models. The logarithm of the distribution function $f$ is plotted as a function of the relative binding energy $\mathcal{E}$. The panels
                                  correspond to different values of the halo to stellar mass ratio, $\mu$. From left
                                  to right and top to bottom, the panels correspond to $\mu = 0,0.1,1,10,10^{2}, 10{^6}$.
                                  For each plot, as shown in the upper-left panel, curves with different colors correspond
                                  to different values of the Sersic index, while different line
                                  types denote different values of the ratio, $x_{_{D}}$, between the effective radii of the halo and stellar components. The $f(\varepsilon)$ has been computed by adopting natural units, where $M_{_L}=1$, $R_{e_L}=1$, and the gravitational constant was set to one.}}
                \label{fig_funzione_di_distribuzione}
\end{figure*}

The distribution function of the $S^{2}$ models is computed by numerical integration of the Eddington formula, as described in App.~\ref{DE2SERSIC}. Fig.~\ref{fig_funzione_di_distribuzione}   plots the $f(\mathcal{E})$   for different values of the free parameters $n$, $\mu$ and  $x_{_{D}}$.  The value of $\mu$ is varied in the range of zero -- no dark matter halo -- to a value of $10^6$, where the stellar component is negligible and the system is completely dark matter dominated. We consider values of $x_{_{D}}$ from $0.1$ to $10^2$, corresponding to the two extreme cases where the dark matter component is either more concentrated or significantly more extended than the luminous one. For
all combinations of $x_{_{D}}$ and $\mu$, different values of $n$ are plotted. We find that for positive values of the relative binding energy the condition $f(\mathcal{E}) \ge 0$ is always fulfilled, implying that the $S^2$ models are physically admissible.

\begin{figure*}[!t]
\centering
    \includegraphics[width=12cm,height=16cm]{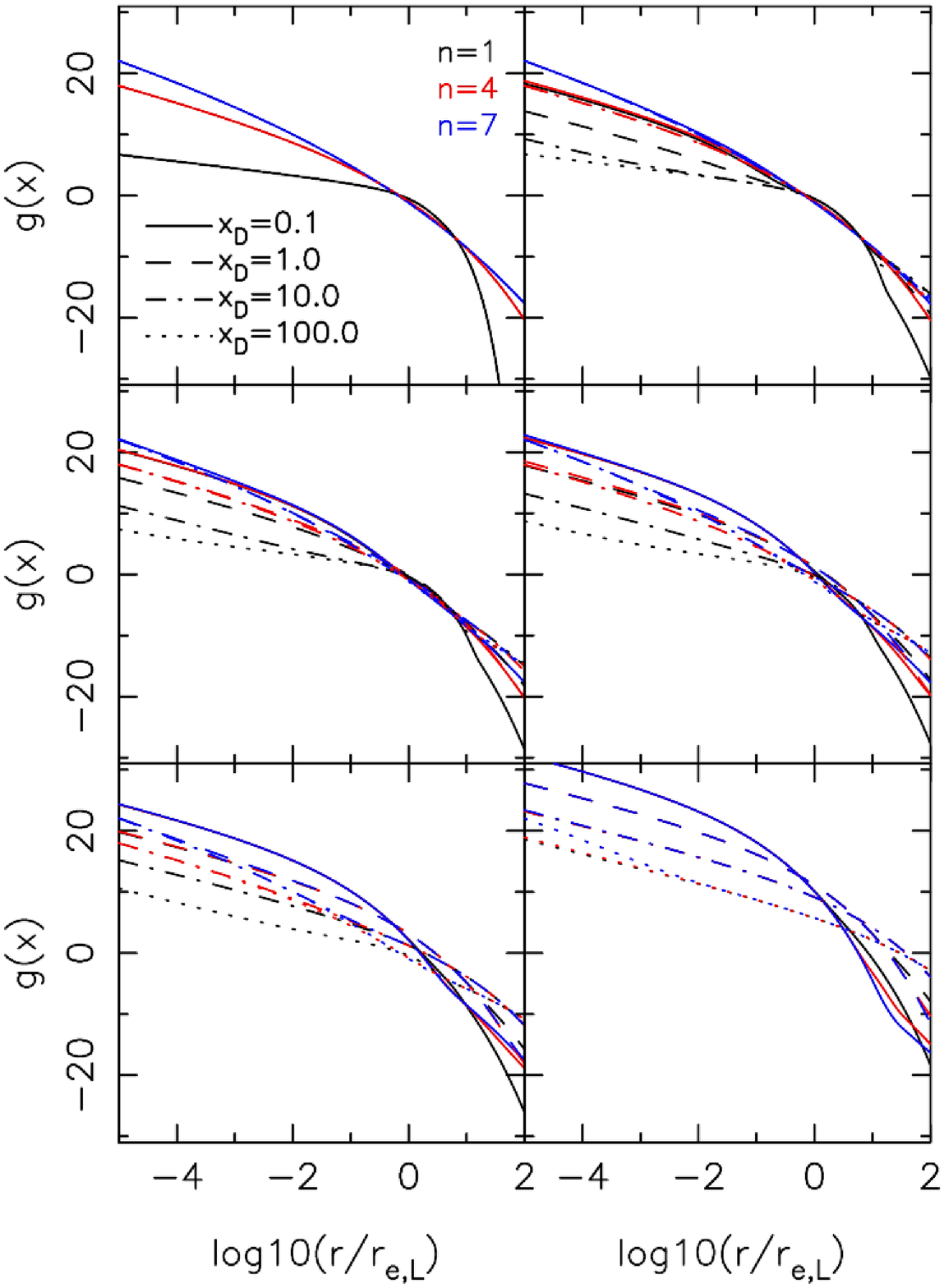}
              \caption{\footnotesize{Stability of the $S^2$ models. The  logarithm of the quantity $g(r)$ (see Eq.~\ref{eq_eddingtopn_positive}) is plotted as a function of the logarithm of the dimensionless radius $r/r_{e,{L}}$. Colors and line types are the same as in Fig.~\ref{fig_funzione_di_distribuzione}. From left
                                  to right and top to bottom, the panels correspond to $\mu = 0,0.1,1,10,10^{2}, 10{^6}$. Natural units have been adopted as for Fig.~\ref{fig_funzione_di_distribuzione}.}}\label{fig_gr}
\end{figure*}

To analyze the stability of the two-component Sersic models, following~\citet{ciotti1991A&A}, we study the sign of the first derivative of the distribution function. According to Antonov’s theorem~\citep[see][pag. 306]{binney&tremaine}, if $\frac{d f}{d \varepsilon} \ge 0$, the system is stable against both radial and non-radial perturbations. As shown in App.~\ref{DE2SERSIC}, a necessary condition for $\frac{d f}{d \varepsilon} \ge 0$ is given by:
\begin{equation}\label{eq_eddingtopn_positive}
    g(r; n, \mu, x_{_{D}}) = - \left [ \frac{d^{2} \rho}{d r^{2}} \left( \frac{d \Psi}{d r}
   \right) - \frac{d \rho}{d r} \frac{d^{2} \Psi}{d r^{2}} \right ] \geq 0 \, .
\end{equation}
For the two-component Sersic models, $g(r)$  is derived numerically as described in App.~\ref{DE2SERSIC}.  Fig.~\ref{fig_gr}   plots $g(r)$  as a function of $r$ for the same sets of $n$, $\mu$, and $x_{_{D}}$ values
as in Fig.~\ref{fig_funzione_di_distribuzione}. The condition $g(r) \ge 0$ is always verified, proving the stability of $S^2$ models.

\section{Physical scales}\label{sec_2_3}

There are five  free parameters that completely characterize the $S^2$
model, i.e. the mass of the stellar component, ${M}_{_{L}}$, its effective radius,  $R_{e_{L}}$,
the Sersic index of the stellar component, $n$, the mass of the dark matter halo, ${M}_{_{D}}$,
and the corresponding effective  radius, $R_{e_{D}}$. Alternatively, one can use the dimensional quantities, ${M}_{_{L}}$ and $R_{e_L}$, and the dimension-less parameters $x_{_{D}}$, $\mu$, and $n$ defined in Sec.~\ref{submodel.2}.
Here, we describe some recipes to express all  the free parameters as a  function
of one single quantity, the absolute luminosity of the stellar
component. This procedure is intended as an handy tool to use the $S^2$ models in merging simulations of ETGs.
We refer to absolute magnitudes in the B band, $M_{B}$, since most of the relations
we use in the following are expressed in that band. In the following, magnitudes are expressed with respect to the Vega system.

The quantity $R_{e_{L}}$ is related to the total luminosity by the Kormendy relation~\citep{kormendy1977ApJ, capaccioli1992}:
\begin{equation}\label{eq_kormendy}
    \log R_{e_{L,B}}=\alpha \langle \! \mu \! \rangle_{e} + \beta \; .
\end{equation}
where $R_{e_{L,B}}$ is the galaxy effective radius in the B-band and $\langle \! \mu \! \rangle_{e}$ is the mean effective surface brigthness inside $R_{e_{L,B}}$. Expressing $R_{e_{L,B}}$ in units of kpc, one has 
\begin{equation}
\langle \! \mu \! \rangle_{e}= -5 \log( R_{e_L}) - M_{B} + 25 +2.5 \log ( 6^8/(2 \pi) ). 
\end{equation}
As discovered by~\citet{capaccioli1992} and~\citet{graham&guzman2003}, ETGs follow two different trends in the $R_e$--$\langle \! \mu \! \rangle_e$ plane, according to their luminosity. The separation between the two families of {\it bright} and {\it ordinary} ellipticals occurs between $M_{B} = -19$ and $M_{B} =-20$. We adopt here a separation value of $-20$. By a linear fit of the data in figure~9 of~\citet{graham&guzman2003}, we obtain $\alpha \sim 0.35$ and  $\beta \sim -6.75$  for the {\it bright} galaxies  ($M_B<-20$) and  $\alpha=-0.02$   and  $\beta=0.45$   for  the \textit{ordinary} ellipticals ($M_B > -20$). The latter value of $\alpha$ is consistent with that of $0.34 \pm 0.01$ found by~\citet{LAB03}, who showed that the Kormendy relation of bright ETGs does not change significantly with redshift up to redshift $z \sim 0.6$ and that the intrisic scatter of the relation amount to $0.4 \pm 0.03$ in $\langle \! \mu \! \rangle_{e}$ (i.e. $\sim 0.14$ dex in  $R_{e_{L,B}}$).
In order to derive the Near-Infrared effective radius $R_{e_L}$, we use Eq.~\ref{eq_kormendy} to compute $R_{e_{L,B}}$ from  $M_B$, and then transform $R_{e_{L,B}}$ into $R_{e_L}$. To this aim, we consider that ETGs have on average a radial color gradient of about $-0.2$ in $B-K$, and that their internal color gradients are observed not to change significantly with galaxy luminosity (see~\citealt{PVJ90,peletieretal1990AJ}). Following~\citet{SpJ93}, the above value of the color gradient implies that the effective radius of ETGs decreases by $\sim 20 \%$ from $B$ to $K$ band. Thus, we derive $R_{e_{L}}$ from the relation  
\begin{equation}
R_{e_{L}} = 0.8 R_{e_{L,B}} \; .
\label{eq_re}
\end{equation}
The Sersic parameter, $n$, of the stellar component depends on luminosity through the magnitude-Sersic index relation~\citep{caonetal1993MNRAS}. \citet{2004ApJ...600L..39T} presented this relation for a sample of $200$ ellipticals at redshift $z \sim 0$. A linear fit to the data in their figure~1 gives~\footnote{We estimate the scatter of the luminosity--Sersic index relation from the distribution of points in Fig.~1 (right--panel) of \citet{2004ApJ...600L..39T}. Assuming that, for a given magnitude, the smallest and largest Sersic index values mark the lower and upper $2 \sigma$ limits around the mean relation, we obtain a $1 \sigma$ dispersion of around $30 \%$ in $n_{B}$ at a given luminosity.}
\begin{equation}
    \log n_B = -0.1219 \cdot M_{B} - 1.6829,
\label{sersiceq}
\end{equation} 
where $n_B$ is the Sersic index of ETGs in the B-band. The Sersic index is not expected to change significantly from optical to NIR wavebands. For instance, as found by~\citet{LAB08}, ETGs have on average  $\log (n_r/ n_K) = -0.007 \pm 0.009$, where $n_r$ and $n_K$ denote the r- and K-band Sersic indices.
Hence, we set $n=n_B$, and use Eq.~\ref{sersiceq} to derive also the NIR Sersic index of the stellar component. 

To express  $R_{e_{D}}$ as a function of $M_{B}$, we use the finding that dark matter halos follow a relation between the half--mass radius, $R_{e_{D}}$, and the average projected surface mass-density inside that radius, $\langle \mu \rangle_{e_{D}}$, similar to the Kormendy  relation of galaxies (\citealt{graham2006b}; hereafter GMM06). This result was obtained from GMM06 for a sample of galaxy-sized dark-matter halos as massive as $10^{12} M_{\odot}$. 
We note that GMM06 derived the quantities $R_{e_{D}}$ and $\langle \mu \rangle_{e_{D}}$ by fitting the projected halo density profile with the Prugniel-Simien  model~\citep{prugniel1997}, i.e. the same kind of profile as adopted here for the dark matter component of the $S^2$ models~\footnote{We notice that GMM06 fitted the Prugniel-Simien model by treating the Sersic index as a free fitting parameter.  Since we fix $n=3$ for the dark-matter halo, the coefficients of Eq.~\ref{GMM06eq}, taken from GMM06, might not be appropriate for our model calibration. When fitting a Sersic model with $n=3$ to a Sersic profile with $n=3.6$ (the average value found by GMM06), we find that the best-fitting effective radius is $\sim 20 \%$ smaller than the true value. However, due to the well-known correlation between effective radius and mean surface brightness, this change in $R_e$ corresponds to a change in $\langle \! \mu \!\rangle_e$, such that points are moved almost parallel to the Kormendy relation ~\citep{LAB03}.}.  We write
\begin{equation}
    \log R_{e_{D}}=\delta \cdot \langle \mu \rangle_{e_{D}} + \zeta.
\label{GMM06eq}
\end{equation}
For systems more massive than $10^{10}$~$M_{\odot}$, GMM06 report a slope of  $\delta \sim 1/3$. This mass range corresponds to $\log R_{e_{D}} > 0.4$ (see fig.~1b of GMM06). Performing a linear fit to the data in figure 2a of GMM06, we obtain $\zeta  \sim 10/3$, with $R_{e_{D}}$ being expressed in units of kpc. 

\begin{figure*}[!t]
\centering
    \includegraphics[width=12cm,height=12cm]{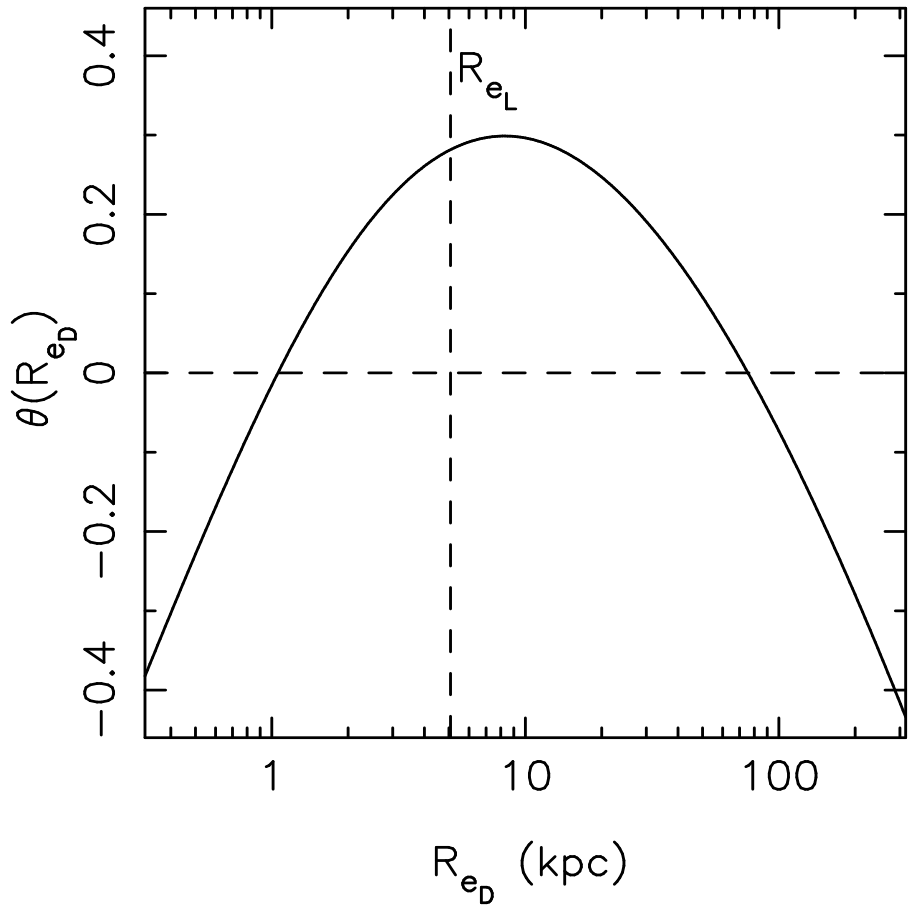}
              \caption{\footnotesize{Derivation of the effective radius of the dark matter component for a galaxy with $M_B=-21$. The $R_{e_D}$ is derived by solving the equation $\theta(R_{e_D})=0$ (Eq.~\ref{eq_solve}). The horizontal dashed line marks the value of $\theta(R_{e_D})=0$, while the vertical dashed line shows the effective radius $R_{e_L}$ of the stellar component. The points of intersection between the horizontal line and the curve denote the values of $R_{e_D}$ which are consistent with our procedure. We consider only the {\it large-halo} solution (right part of the plot), with $R_{e_D} \! > \! R_{e_L}$ (see the text).}}\label{solve}
\end{figure*}

Then, we derive the mass of the dark-matter and stellar components as a function of the B-band magnitude, using the recent results obtained from~\citet{cappellarietal2006MNRAS} (hereafter CAP06) for elliptical    and    lenticular    galaxies    in   the SAURON    project~\citep{baconetal2001}. From the relation between dynamical  mass-to-light ratio in  $I-$band and total mass of CAP06 (see their eq.~9), one obtains:
\begin{equation}\label{eq_total_mass}
    {M}_{e_{L}}+{M}_{e_{D}} = 1.175 \cdot 10^{0.1317 - 0.528 \cdot M_B} \; .
\label{CAP06eq}
\end{equation}
where ${M}_{e_{L}}$ and ${M}_{e_{D}}$  denote the masses of the stellar and dark matter components within $R_{e_L}$. This relation provides the total dynamical mass with an accuracy of $\sim 30 \%$. Following~\citet{fukugita1995PASP}, we derive Eq.~\ref{CAP06eq} by assuming a typical $B \! - \!I$ color term~\footnote{We notice that the assumption of a constant color term for ETGs is just a simplified assumption, since early-type systems are known to follow a color--magnitude relation (e.g.~\citealt{VS77}). Though the above procedure can be generalized to account for a given color--magnitude relation, we decided to fix $B-I$. In fact, one should notice that the slope of the color-magnitune relation might be significantly affected from the aperture where color indices are derived, due to the existence of internal color gradients in galaxies~\citep{SCO01}, with the slope flattening more and more as larger apertures are adopted.} for elliptical galaxies of  $2.23$ and  the B- and I-band magnitudes of the Sun to be  $5.51$ and $4.08$, respectively. Under the assumption of a radially constant $M_{_L}/L$ ratio, one has ${M}_{_{L}}=2{M}_{e_{L}}$. According to CAP06, ${M}_{e_{L}}$ is about $0.16$dex smaller~\footnote{This result was obtained under the assumption of a Kroupa IMF.} than the dynamical mass within $R_{e_L}$, i.e. ${M}_{e_{L}} \sim 0.6918 \left( {M}_{e_{L}}+{M}_{e_{D}} \right)$. Thus, from Eq.~\ref{CAP06eq}, one obtains:
\begin{equation}\label{eq_massl}
    {M}_{e_{L}}= 0.81286 \cdot 10^{0.1317 - 0.528 \cdot M_B} \; ,
\end{equation}
and
\begin{equation}\label{eq_massd}
    {M}_{e_{D}}= 0.36214 \cdot 10^{0.1317 - 0.528 \cdot M_B} \; .
\end{equation}
In order to relate ${M}_{e_{D}}$ to ${M}_{_{D}}$, we use the analytic expression for the projected luminosity profile of the Sersic model (see eq.~2 of~\citealt{ciotti1999}). Since the dark-matter component is described by a Sersic model having $n=3$, we can write:
\begin{equation}
M_{e_{D}} = M_{_D} \cdot \gamma \left( 6, b_3 \cdot \left( \frac{R_{e_L}}{R_{e_D}}\right)^{1/3} 
\right)
\label{eq_md_mde}
\end{equation}
where $\gamma$ denotes the normalized incomplete gamma function~\footnote{The normalization of the incomplete gamma function is done by dividing it with the complete gamma function. We notice that in Sec.~2.1, we adopt a different notation where the $\gamma$ function is not normalized.}, and $b_3 = 5.6631$. The quantity $b_3$ is computed by setting $n=3$ in the analytic approximation of $b$ reported in Sec.~\ref{submodel.1}.
From Eqs.~\ref{eq_massl} and Eq.~\ref{eq_massd}, one obtains:
\begin{equation}
M_{_D} = \frac{ 0.36214 \cdot 10^{0.1317 - 0.528 \cdot M_B} }{\gamma \left( 6, b_3 \cdot \left( \frac{R_{e_L}}{R_{e_D}}\right)^{1/3} \right)} \; .
\label{eq_md}
\end{equation}
In practice, for a given $M_B$, $R_{e_L}$ is computed from Eqs.~\ref{eq_kormendy} and~\ref{eq_re}, and the quantities $R_{e_D}$ and $M_{_D}$ are derived by solving simultaneouly Eqs.~\ref{eq_md} and~\ref{GMM06eq}. This is equivalent to solve the non-linear equation
\begin{equation}
7.1077+\frac{\zeta}{2.5\delta}+0.528 M_B+\log\left[ \gamma \left( 6, b_3 \cdot \left( \frac{R_{e_L}}{R_{e_D}}\right)^{1/3} \right) \right]  + \frac{5 \delta - 1}{2.5 \delta} \log(R_{e_D})=0
\label{eq_solve}
\end{equation}
with respect to $R_{e_D}$. We denote the first member of this equation as $\theta(R_{e_D})$.
As an example, Fig.~\ref{solve} plots $\theta(R_{e_D})$ as a function of $R_{e_D}$, for the case $M_B=-21$. The figure shows that Eq.~\ref{eq_solve} has in general two distinct solutions, corresponding to the points where the horizontal dashed line in the figure crosses the curve. One has a {\it small-halo} solution with $R_{e_D}<R_{e_L}$ (and $M_{e_D}<M_{e_L}$), and a {\it large-halo} case, whereby the  dark-matter component is larger and more massive than the 
stellar one. In the small-halo case, the $M_{e_D}$ value is four (eight) times
smaller than $M_{e_L}$ for $M_B=-22$ ($-20$), while $R_{e_D}$ is three (ten) times smaller
than $R_{e_L}$. This would imply that almost all the dark matter in ETGs should be enclosed within one $R_{e_L}$, in disagreement with dynamical, X-Ray, and  weak lensing studies~\citep{matsushita98, wilson01, Ger01}. Therefore, we consider here only the large-halo 
solutions of Eq.~\ref{eq_solve}. We notice that Eq.~\ref{GMM06eq} applies to the case of 
massive galaxy-sized halos ($M_{_{D}} \sim 10^{12}$), which might be appropriate only for {\it bright} galaxies ($M_B < -20$). For galaxies fainter than $M_B = -20$, we fix~\footnote{Applying Eqs.~\ref{GMM06eq} and~\ref{eq_solve} also for $M_B > -20$ would lead to an improbable set of solutions where systems fainter than $M_B=-18$ would have dark-matter halos more massive than a galaxy with $M_B=-22$. } the ratio of   dark to stellar effective radius to the value obtained for $M_B = -20$ and then derive the total dark-matter mass from Eq.~\ref{eq_md}.

To  summarize, we use  the Kormendy  and the  luminosity--Sersic index
relations to express $R_{e_{L}}$ and $n$ as a function of $M_B$. Then,
by  using Eq.~\ref{eq_massl} and  solving Eq.~\ref{eq_solve},  we also
express ${M}_{_{D}}$,  $R_{e_{D}}$, and ${M}_{_{L}}$ as  a function of
$M_B$.   In Tab.~\ref{tab_par_galassie},  as an  example, we  show the
values  of the  five  free parameters  of  the $S^2$  models that  are
obtained from  the above procedure  in six cases equally  spanning the
magnitude range of $-22$ to  $-17$. In general, the procedure leads to
have galaxy models where the dark matter component is less massive and
less extended  in lower luminosity systems. On the other hand, the relative amount  of dark
matter  within $R_{e_L}$  does  not depend  on  galaxy luminosity,  in
agreement   with  the  finding   of  CAP06   (see  Eqs.~\ref{eq_massl}
and~\ref{eq_massd} above).

We remark that the above procedure derives the free parameters of the $S^2$ models
by using the observed properties of early-type systems at $z \sim 0$.
Hence, one possible caveat when applying the above procedure to merging simulations is 
that such properties might not necessarly be the same for the high-redshift 
progenitors of ETGs. Moreover, one should consider that most of the observed relations 
(such as the Kormendy and the luminosity-size relations) of ETGs have significant intrinsic dispersion (see the values reported above), implying a dispersion, at a given magnitude, also in 
the parameter's values reported in  Tab.~\ref{tab_par_galassie}.

\begin{table*}[!t]
\begin{center}
\begin{tabular}{|c|c|c|c|c|c|c|}  \hline \hline
 $M_{_{B}}$  &   ${M}_{_{L}}$  &   $R_{e_{L}}$  &   $ {M}_{_{D}}$     &   $R_{e_{D}}$ & $n$ & $ {M}_{e_{D}}$   \\
   &  ($ 10^{10}$~$M_{\odot}$)   &   (kpc)   &   ($10^{10}$~$M_{\odot}$)    &   (kpc) &  &  $(10^{10}$~$M_{\odot}$) \\
(1) & (2) & (3) & (4) & (5) & (6) & (7)\\
  \hline  \hline
-22 &      90.94 &      14.84 &     267.20 &     108.60 &     10.0 &      20.25 \\
-21 &      26.96 &       5.07 &     199.74 &      75.50 &      7.5 &       6.00 \\
-20 &       7.99 &       1.73 &     133.26 &      45.50 &      5.7 &       1.78 \\
-19 &       2.37 &       0.90 &      39.53 &      23.79 &      4.3 &       0.53 \\
-18 &       0.70 &       0.87 &      11.72 &      22.81 &      3.2 &       0.16 \\
-17 &       0.21 &       0.83 &       3.47 &      21.88 &      2.5 &       0.05 \\
\hline  \hline
\end{tabular}
\caption{\footnotesize{Derivation of the free parameters of the double Sersic models as a function of luminosity. The columns are: (1) $B-$band magnitude, $M_{B}$, of the stellar
 component, (2) total stellar mass, ${M}_{_{L}}$, (3) effective radius of the stellar component, $R_{e_{L}}$, (4) total mass of the dark matter halo, ${M}_{_{D}}$, (5) effective radius of the dark matter component, $R_{e_{D}}$, (6) Sersic index $n$, and (7) mass of the dark matter halo within $R_{e_{L}}$.}}
\label{tab_par_galassie}
\end{center}
\end{table*}

\section{Optimal softening length}\label{sec_2_4}

Performing discrete realizations of galaxy models requires that
a given gravitational {\it softening} parameter, $\epsilon$, is adopted. The value of $\epsilon$ should depend on the number of particles, $N$, defining the mass  and spatial
resolution of the simulation. Here, we discuss how to set $\epsilon$ and $N$
for the Sersic models.

\subsection{The optimal smoothing length}

\begin{figure}
\begin{center}
\includegraphics[width=8cm]{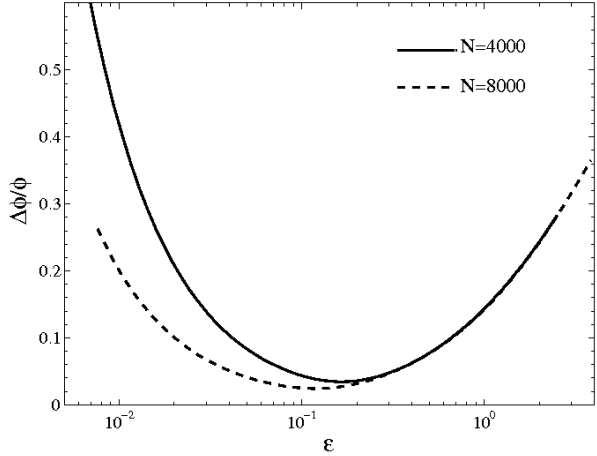}
\caption{\footnotesize{Mean value of the relative error on the gravitational potential as a function of the softenin length, $\epsilon$, for a Sersic model with $n=4$. As shown in the upper-right corner of the plot, the dashed and solid curves correspond to discrete realizations with a different number of particles, $N$. }}
\label{deltaphi}
\end{center}
\end{figure}

\begin{figure}
\centering
\includegraphics[width=8cm]{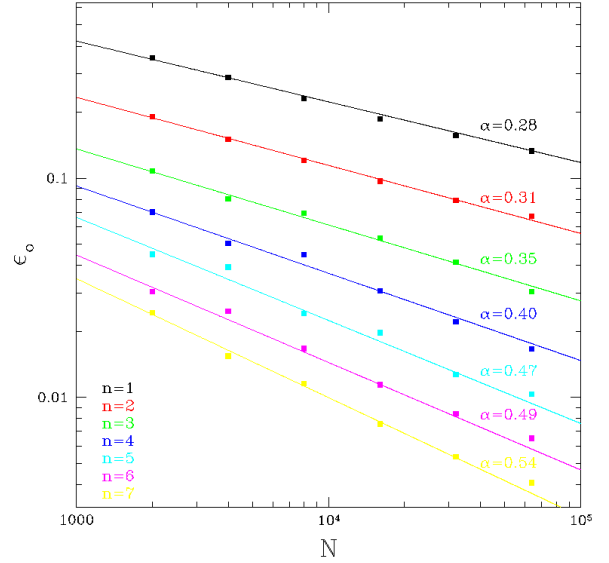}
\caption{\footnotesize{Dependence  of  the  optimal  softening length of one-component models,  $\epsilon_0$,  on the number of particles, $N$, for different values of the Sersic index, $n$. Different colors correspond to different values of $n$ as shown in the lower-left corner of the plot. Solid lines plot the best-fitted power laws to the trends of $\epsilon_o$ vs. $N$ (see the text). The exponent $\alpha$ of each fitted power-law is reported on the top--right of the corresponding line.}}
\label{fig_softening}
\end{figure}

\begin{figure}
\begin{center}
\includegraphics[width=8cm]{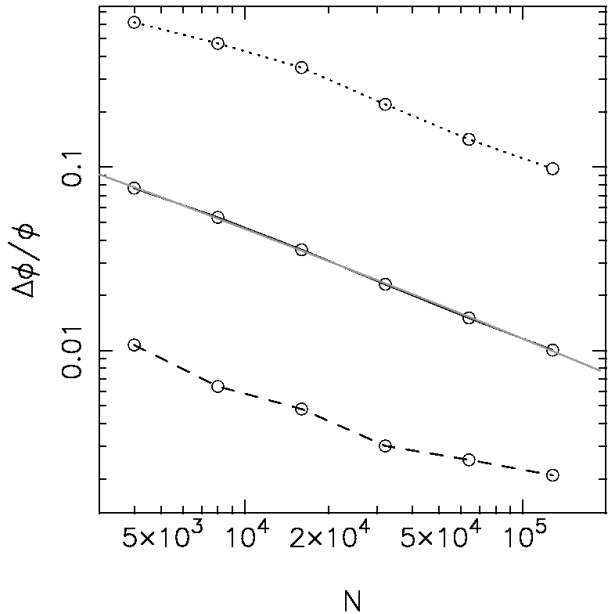}
\caption{\footnotesize{The relative error on the gravitational potential, $\Delta \phi/\phi$, is plotted as a function of the number of particles, $N$, for three one-component Sersic models having $n=1$ (dashed line), $n=4$ (solid line) and $n=7$ (dotted line), respectively. The gray line shows the power-law fit, $\Delta \phi/\phi \propto N^{-1/2}$, to the points for $n=4$.}}
\label{eps_n}
\end{center}
\end{figure}

Usually, the  value of $\epsilon$ is  chosen with some \textit{ad hoc}
prescription. One fixes the total   number   of   particles in the simulation
(which is limited from the available CPU resources) and then assigns the $\epsilon$
in order to achieve the desired spatial resolution.
\cite{merritt1996} (hereafter MER96) showed that the softening length
of  an  N-body  system  can  be  chosen in  an  objective (\textit{optimum})
way  by minimizing the average  error in the gravitational force computation
over the whole space. Following  a  similar   approach,  we  assign $\epsilon$  by
minimizing the  average error in the computation  of the gravitational
potential. We  consider here the spline softening kernel of~\citet{ML85},
which is implemented into the simulation code Gadget-2~\citep{springel05}.
Hereafter, we express $\epsilon$ in units of the effective radius, $R_{e_L}$.\\
We start by considering the case of single Sersic models. For a given Sersic index, $n$, and a given number of particles,  $N$, we generate several realizations of
the deprojected Sersic  model.
For a given realization,  we calculate the
softened gravitational potential at the position of each particle and
the corresponding true gravitational potential (Eq.~\ref{eq_potential_definitiva}). Then, the rms of the relative absolute differences between the softened and true potential, $\Delta \phi/\phi$, is computed over all the particles. We average the value of $\Delta \phi/\phi$ over 100 realizations. Fig.~\ref{deltaphi} shows  how the mean value of $\Delta \phi/\phi$ changes as a function of $\epsilon$.  As example, the figure plots the case of a de Vaucouleurs model ($n=4$) for two different values of $N$.
In both cases, there is a minimum in $\Delta \phi/\phi$. Following an argument similar to that of MER96, the existence of a minimum can be explained as follows. For low $\epsilon$, the error is dominated by the differences between the point-like Newtonian potential of each particle and  the true gravitational potential.  Increasing $\epsilon$, these differences become smaller and $\Delta \phi/\phi$ decreases.  For large $\epsilon$, the discrete potential is smoothed on a scale larger than the typical interparticle separation~\footnote{The softening mostly affects the region where the potential changes more rapidly, i.e. the region inside the effective radius $R_{e_{L}}$. With typical interparticle separation, we refer to some statistical estimator of the average particle-particle distance  within that region, such as the mode or the median of the distribution of interparticle distances.} and the discrete potential is overly smoothed with respect to the true gravitational potential. Increasing $\epsilon$, this large-scale smoothing becomes more and more important, and the value of $\Delta \phi/\phi$ increases as well. For a given number of particles,  we  define the position of the minimum as the optimal smoothing length, $\epsilon_{o}$. Increasing the number of particles, the typical interparticle separation,  $d_N$, decreases, and thus the optimal smoothing is obtained for smaller $\epsilon$. Fig.~\ref{fig_softening}  plots $\epsilon_{o}$   as  a  function of $N$ for different  values of  the Sersic index. The optimal  softening  length  turns out  to  decrease as  either $n$  or $N$ increase. This is due to the fact that, in both cases, the typical
particle separation,  $d_N$, decreases. In particular,  when $n$  increases, the  mass profile of the model is  more concentrated in the center and,  at fixed $N$, $d_N$  is smaller.  As shown in Fig.~\ref{fig_softening}, the  trend of $\epsilon_o$ vs.  $N$ can be accurately modeled by
a power law, $\epsilon_o = \beta N^{-\alpha}$, where both $\alpha$ and $\beta$ depend on the value of $n$. The value of  $\alpha$ changes from $\sim 0.28$ for $n=1$ to
$\sim 0.54$ for $n=7$. For $n \le 2$, the shape of the Sersic profile is flatter than for higher values of $n$, and the $\epsilon_o$ is essentially proportional to the mean interparticle separation, with $\epsilon_o \propto N^{-1/3}$. For a de Vaucouleurs profile ($n=4$), the value of $\alpha$ is $\sim 0.4$, in agreement with that of $0.44$ found by MER96 for the Hernquist model.  For higher $n$, the Sersic profile becomes more and more peaked in the center and the value of $\alpha$ deviates more and more from the simple $\alpha \sim 1/3$ expectation.
Fig.~\ref{eps_n} shows how the mean relative error on the potential, $\Delta \phi/ \phi$, depends on the number of particles and the Sersic index when adopting the optimal smoothing parameter. For a given Sersic model, the error decreases with $N$ following the power-law $\Delta \phi/ \phi \propto N^{-1/2}$, in agreement with what found by MER96 for the Hernquist model. For a given $N$, the error is larger for higher Sersic index. Hence, if a given accuracy in the computation of the gravitational potential has to be achieved, for higher $n$ a larger number of particles has to be adopted.

\subsection{Models in isolation}
\label{mod_is}
\begin{figure}[h]
 \centering
        \includegraphics[width=8cm]{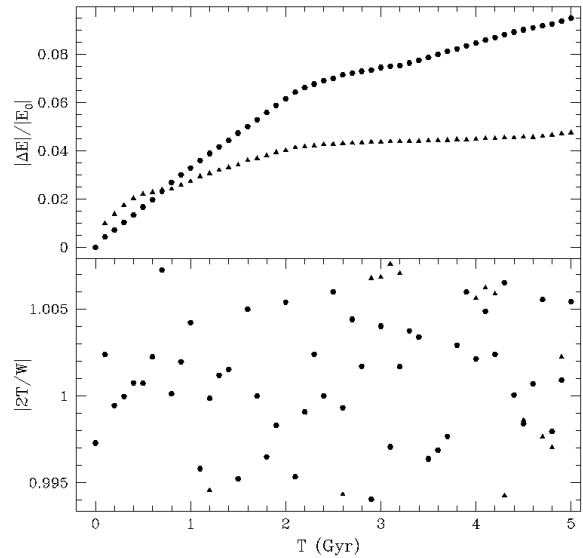}
           \caption{\footnotesize{Variations in total energy and virial ratio as a function of time. Triangles and circles correspond to one- and two-component models, respectively  (see the text). Notice that the deviations from conservation of total energy ($|\Delta E/E_0|=0$) and the virial equilibrium ($ |2 T / W |=1$) are small for both models.}}\label{plot_energy}
\end{figure}

\begin{figure}[t]
\centering
         \includegraphics[width=8.5cm]{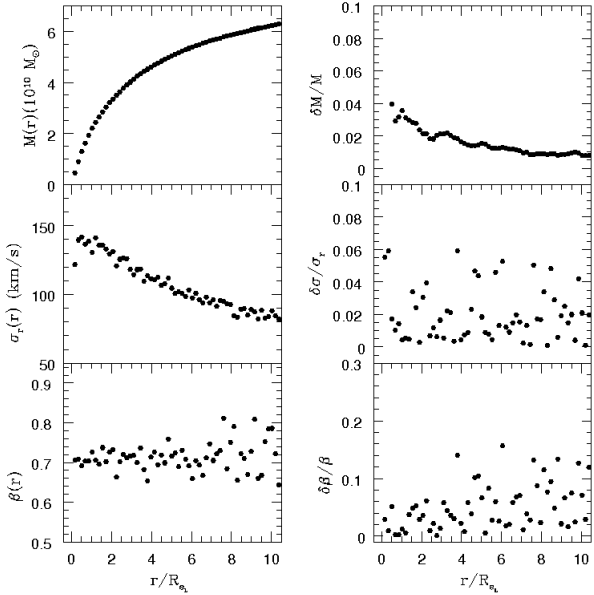}
                \caption{\footnotesize{Evolution in mass, velocity dispersion, and anisotropy profiles of single Sersic models. We plot the case of the model (1) described in the text. Left panels plot the mass (top), velocity dispersion (middle) and anisotropy (bottom) profiles  of the model at $T=0$~Gyr. The right panels show the relative absolute radial variation of the profiles after $T=5$~Gyrs. For each value of the spatial radius $r$, the variation is computed with respect to the initial value at that radius. We note that the variations from $|2T/W|=1$ are small for both models.}}\label{fig_profile_lum_only}
\end{figure}

\begin{figure}[t]
\centering
         \includegraphics[width=8.5cm]{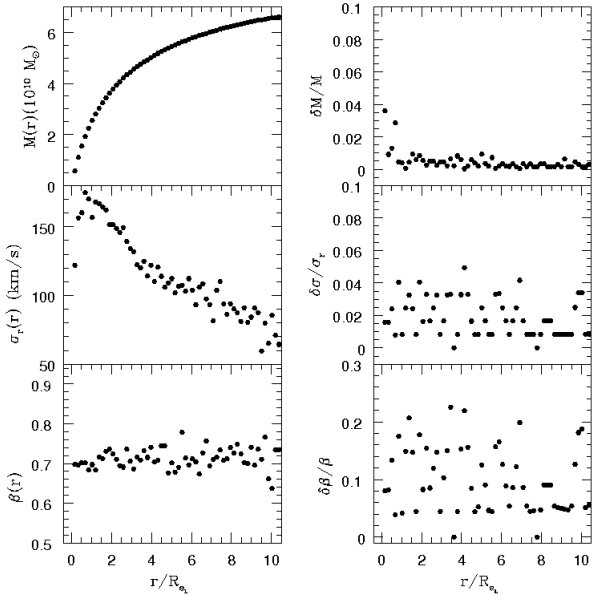}
                \caption{\footnotesize{Same as Fig.~\ref{fig_profile_lum_only} for the luminous component of $S^2$ models. The case of model (2) is shown (see the text). }}\label{fig_profile_lum}
\end{figure}

To perform discrete realizations of the $S^2$ models, one can
adopt different softening lengths for the stellar and dark matter components, according to the optimal definition given above. However,
these softening parameters represent an optimal choice only for
one-component Sersic models, and we are not guaranteed that they provide
also an accurate choice for the two-component models.
To verify that the optimal prescription for $\epsilon$ gives sensible results
even in the case of two-component models, we compared the evolution of
double and single Sersic models in isolation. As example, we consider here (1)
a one-component model with ${M}_{_{L}} \sim 27 \cdot 10^{10}$~$M_{\odot}$ and
$R_{e_{L}}\sim5$~kpc, and (2) an $S^2$ model whose parameters are the same as those reported in Tab.~\ref{tab_par_galassie} for the case $M_B=-21$.  Model (1) is obtained by considering only the stellar component of model (2). To evolve the models in isolation,
we adopt  $50000$ particles of luminous matter in both cases and $75000$ particles   of dark matter for model (2).  Looking at Fig.~\ref{eps_n}, we see that adopting the optimal smoothing parameter for these values of $N$ allows an accuracy better than $10 \%$ on the gravitational potential to be achieved.
The simulations were ran over $5$~Gyrs with the simulation code Gadget-2, using a Beowulf system with thirty-two AMD-Opteron 244 processors. As initial conditions, we created discrete realizations of the models by computing their  density profile and distribution function with the set of Fortran codes that are made publicly available (see App.~\ref{code}). The softening parameters were chosen according to Fig.~\ref{fig_softening}. For the stellar component, we adopt $\epsilon_o=0.013$~kpc, while for the dark matter component we set $\epsilon_o=0.053$~kpc. \\
Fig.~\ref{plot_energy} (upper panel) plots the relative absolute variation of the total  energy of both systems, $|\Delta E / E_0|$,  as a function of time, where $E_0$ is the total initial energy of the simulation.  Apart from a small and slow  secular  drift, one can see that for both models the total energy of  the system is  preserved, with a value of $|\Delta E / E_0|$ smaller than $\sim 8 \%$ after $5$~Gyrs. Fig.~\ref{plot_energy} (lower panel) also shows the  evolution of  the virial
ratio, $|2T/W|$,  where $T$  and $W$ are  the total kinetic  and potential energy of the system, as a function of time. For both the single and $S^2$ models, the deviations from  the virial equilibrium, $2T/W=1$, are small,  amounting to at most $\sim 0.7\%$ in modulus after $5$~Gyrs.  \\
Fig.~\ref{fig_profile_lum_only} plots, for the one-component model,  the radial profiles in mass, velocity dispersion, and anisotropy at $T=0$~Gyrs (left panels), and the relative variations of these profiles after the model has been evolved for $5$~Gyrs (right panels). The profiles are plotted in a radial range of $r_{min}=3 \epsilon_o$ to $r_{max}=5R_{e_{L}}$. The value of $r_{min}$ is chosen in order to avoid the inner region of the model which is affected by the smoothing in the gravitational potential. The maximum radius, $r_{max}$, is set to a sensible value where one can compare the model to the observed profiles of ETGs. The simulation shows that the profile in mass is preserved within a few percentages over the whole radial extent. The velocity dispersion and the anisotropy profile are also preserved within $\sim 10\%$.
Fig.~\ref{fig_profile_lum} plots the same profiles as in Fig.~\ref{fig_profile_lum_only} for the stellar component of model (2). Remarkably, all the profiles are preserved even in this case within $\sim 10 \%$ over at least $5$~Gyrs. The same result was obtained when considering the properties of the dark-matter component of model (2), and for all the $S^2$ models  whose parameters are
listed in Tab.~\ref{tab_par_galassie}.

\section{Summary and Discussion}\label{sec_2_7}
We have presented models of ETGs consisting of a stellar component and a dark matter halo that follow the deprojected Sersic law. The models describe non-rotating, isotropic, spherical systems, whose  density--potential pair is derived  under the assumption that the stellar mass-to-light ($M_{_L}/L$) ratio of galaxies does not depend on radius. 

As mentioned in Sec.~\ref{submodel.1}, the constant $M_{_L}/L$ assumption might not reflect the real physical properties of ETGs. Galaxies are observed to have internal color gradients, reflecting variations of stellar population properties (such as age and metallicity) from the galaxy center to the outskirts (e.g.~\citealt{PVJ90}). It has been shown that (i) color gradients are mainly driven by a mean metallicity gradient in the range of $\nabla_Z = -0.2$ to $\nabla_Z=-0.3$, with an uncertainty of $\sim 0.1$; and that (ii) a small positive age gradient of $\nabla_t \sim 0.1$ is also consistent with observations (see e.g.~\citealt{peletieretal1990AJ, SMG00, IMP02,  LBM03,  TaO03}). Here, we denote as $\nabla_Z$ and $\nabla_t$ the logarithmic variations of metallicity and age per decade in galaxy radius. From the theoretical viewpoint, age gradients are expected to arise in the formation of ETGs by gas-rich mergers, where early-type remnants are better described by a two-component stellar profile, with the two components having different ages~\citep{hopkins2008a}.   We can use the above values of $\nabla_Z$ and $\nabla_t$ to infer the corresponding radial variations of  $M_{_L}/L$. Using single stellar populations models from~\citet{BrC03} with a Scalo IMF and an age of $12$Gyr~\footnote{In a cosmology with $\Omega_{\rm m}$ = 0.3,  $\Omega_{\Lambda}$ = 0.7, and $\rm H_{\circ}$ = 70 km $\rm s^{-1}$ $\rm Mpc^{-1}$, this would correspond to a formation redshift of $z \sim 4$.}, one obtains that a metallicity gradient of $\nabla_Z = -0.2$ ($-0.3$) corresponds to a variation of $34 \%$ ($51 \%$) in the B-band  $M_{_L}/L$ per decade of galaxy radius. This variation largely decreases in K-band, where the inferred variation of $M_{_L}/L$ amounts to $\sim 15 \%$ ($24 \%)$. Considering a positive age gradient of $0.1$dex, the $M_{_L}/L$  variation would further decrease to about $7\%$ ($10 \%$) in K-band, while the above uncertainty on color gradients would translate to an error of about one third in the estimated $M_{_L}/L$ percentages.  We conclude that, provided one adopts the K-band light profile of ETGs to infer the underlying distribution of stellar matter, the assumption of a constant $M_{_L}/L$ is empirically well motivated. 

For what  concerns the other assumptions underlying the $S^2$ models, one should notice that ETGs actually  span  a wider range of kinematical and structural properties than that considered here. 
For instance, the $S^2$ models populate the origin of the anisotropy ($v/\sigma$ vs. ellipticity) diagram,  while  ETGs  populate different regions of it.  In order to explore  the  corresponding effect  on dry-merging simulations,  some studies have realized merging simulations  where the progenitors are obtained by either dissipationless~\citep{NKB06} or dissipational~\citep{CDD06, RBC06} merging of 
disk systems. This re-merger approach has the main advantage that progenitors span a wide range of 
ETG properties, such as $v/\sigma$, ellipticty, and isophotal shape. Though neglecting these aspects,
the $S^2$ models have the main advantage of allowing one to explore a key  observational  feature: 
the  wide range of profile shapes observed in early-type systems~\citep{caonetal1993MNRAS}. Moreover, re-merging of $S^2$ would likely allow one to further enlarge the range of kinematic and isophotal properties of merging progenitors.

The free parameters of the two components of $S^2$ models are assigned in order to match the observed properties of ETGs as well as recent results of N-body simulations of galaxy-sized dark matter halos. We report a concise reference to
the basic integral equations that define the density-potential pair and the distribution function of the deprojected  Sersic law, showing how these equations can be used to define the $S^2$ models. We show that for all possible values of the free parameters of the models, the total distribution function is always non-negative defined, implying that the models
are physically admissible solutions of the collisionless Boltzmann equation. Moreover, the first derivative of the total distribution function is always non-negative defined, implying that the models are stable against radial and non-radial perturbations.
For a given Sersic model, we present an objective prescription to adopt an optimal smoothing length of discrete model realizations. The optimal smoothing length is defined as the softening parameter that minimizes the error on the gravitational potential of the system, and depends on the Sersic index $n$ as well as on the number of particles of the simulation. The power-law relations that describe these trends are reported, with the aim of providing a prescription to create discrete realizations of $S^2$ systems, whose discrete gravitational potential closely matches the true model potential. As a caveat, when using such a prescription for merging simulations, one should notice that the optimal smoothing length for the progenitors might not necessarely concide with the optimal softening for the merging remnants, depending on the structural properties (i.e. the Sersic index) of the merging end-products.
This issue can be addresses by exploring the effect of changing the number of particle, and the corresponding smoothing length, of the colliding systems.

We provide the Fortran code that allows one to calculate all the properties of single and double Sersic models. The code together with the recipes for computing the optimal softening scale are intended as general tools to perform merging simulations of early-type galaxies,
whereby the structural non-homology of these systems (i.e. the variation of the shape parameter along
the galaxy sequence) might be taken into account. In a companion contribution (Coppola et al. 2009b, in preparation), we use the $S^2$ models to investigate how dissipation-less (major and minor) mergers affect the structural properties of ETGs, such as the shape of their light profile and their stellar population gradients.

\begin{acknowledgements}
We thank L. Mayer and E. D'Onghia for the helpful comments and suggestions. We also thank the referee
who provided several comments/suggestions which helped us to significantly improve this manuscript.
\end{acknowledgements}
\appendix

\section{Fortran codes}
\label{code}
The properties of both the single and double Sersic models are computed by a set of FORTRAN routines. All the Fortran codes are made publicly available~\footnote{http://www.na.astro.it/$\sim$labarber/Sersic}. For the one-component models, the code allows the user to calculate the density, mass, and gravitational potential profiles (by a numerical integration of Eqs.~\ref{eq_dens_definitiva},~\ref{eq_mass_definitiva}, and~\ref{eq_potential_definitiva}), as well as the distribution function (App.~\ref{DE2SERSIC}). Other quantities, such as the total potential and gravitational energy of the system, its spatial and projected velocity dispersion profiles, are also computed by specific Fortran routines.
For the double Sersic model, since the computation of the density-potential pair is time-demanding, we proceed as follows.
\begin{itemize}
  \item[-] For a given value of the Sersic index $n$, that characterizes the luminous component of the model, we calculate the dimension-less mass, density, potential and the first and second derivatives of the density profile over a grid in the dimension-less spatial radius $x$. The same computation is done for the Sersic index  of the dark matter component, $n=3$ (Sec.~\ref{DF}).
  \item[-] The total density-potential pair and the distribution function are then obtained by interpolating the above radial profiles. To this effect, the values of the parameters $\mu$ and $x_{{D}}$ of the model have to be provided (Sec.~\ref{DF}).
\end{itemize}
The software to perform this interpolation procedure is also provided.

\section{Distribution function of the double Sersic model}
\label{DE2SERSIC}
In order to apply the Eddington inversion (Eq.~\ref{eq_distribution_function}),
one has to calculate the function $\frac{d^{2} \rho}{d \Psi^{2}}$, where $\rho$
is the spatial density profile and $\Psi  \equiv - \varphi + \varphi_0$ is the rescaled gravitational potential (see Sec.~\ref{DF}).
We start from the following identity:
\begin{eqnarray}
    \frac{d^{2} \rho}{d \Psi^{2}} &=& \frac{d^{2} \rho}{d r^{2}}
    \left ( \frac{d \Psi}{d r} \right )^{-2} - \frac{d \rho}{d r}
    \left( \frac{d \Psi}{d r} \right
    )^{-3} \frac{d^{2} \Psi}{dr^{2}}.
\label{d2rd2p}
\end{eqnarray}
Then, using the fact that $\rho(r)=\rho_{_{L}}+\rho_{_{D}}$ and $\varphi(r)=\varphi_{_{L}}+\varphi_{_{D}}$, one obtains the  following expression:
\begin{eqnarray}
\label{eq_eddington_finale}
    \left[ \frac{d^{2} \rho}{d r^{2}} \left ( \frac{d \varphi}{d r} \right
    ) - \left(\frac{d \rho}{d r}\right) \frac{d^{2} \varphi}{dr^{2}}
    \right]  = \left[ \frac{d^{2} \rho_{_{D}}}{d r^{2}} \frac{d \varphi_{_{D}}}{d r} -
    \frac{d \rho_{_{D}}}{d r} \frac{d^{2} \varphi_{_{D}}}{dr^{2}}\right] + \nonumber \\
     \left[\frac{d^{2} \rho_{_{L}}}{d r^{2}} \frac{d \varphi_{_{L}}}{d r}
     -     \frac{d \rho_{_{L}}}{d r} \frac{d^{2} \varphi_{_{L}}}{dr^{2}}\right]
    - \left[ \frac{d \rho_{_{L}}}{d r} \frac{d^{2} \varphi_{_{D}}}{d r^{2}} -
    \frac{d \rho_{_{D}}}{d r} \frac{d^{2} \varphi_{_{L}}}{dr^{2}}\right] \; .
\end{eqnarray}
The first and second derivatives of $\rho_{_{L}}$ and $\rho_{_{D}}$ can be derived by numerically
differentiating Eqs.~\ref{eq_dens_definitiva} and~\ref{eqs_DM1}. The derivatives of the gravitational potential and the density profile can be obtained from the expression of the gravitational potential and the mass profile  of the stellar and dark matter components, using the following identities:
\begin{equation}
    \frac{d \varphi_{_{L}}}{dr} = \frac{G {M}_{_{L}}}{R_{e_{L}}^{2}}
    \frac{\widetilde{M}(x)}{x^{2}}|_{x=r/R_{e_{L}}} \; ,
\label{d1pl}
\end{equation}
\begin{equation}
    \frac{d \varphi_{_{D}}}{dr} = \frac{G {M}_{_{L}}}{R_{e_{L}}^{2}} \frac{\mu}{x_{_{D}}^{2}}
    \frac{\widetilde{M}(x/x_{_{D}})}{(x/x_{_{D}})^{2}}|_{x=r/R_{e_{L}}} \; ,
\label{d1pd}
\end{equation}
\begin{equation}
    \frac{d^{2} \varphi_{_{L}}}{dr^{2}} = 4 \pi G \frac{{M}_{_{L}}}{R_{e_{L}}^{3}}
    \frac{b^{2n}}{2 \pi n \Gamma(2n)} \widetilde{\varphi}_{_{L}}(x) - \frac{2}{R_{e_{L}}}
    \frac{1}{x} \frac{d \varphi_{_{L}}}{dr} \; ,
\end{equation}
\begin{equation}
    \frac{d^{2} \varphi_{_{D}}}{dr^{2}} = G \frac{{M}_{_{L}}}{R_{e_{L}}^{3}}
    \frac{\mu}{x_{_{D}}^{3}} \left [ \frac{2 b^{2m}}{m \Gamma(2m)}
    \widetilde{\varphi}_{D}(x/x_{_{D}}) - \frac{2 M(x/x_{_{D}})}{(x/x_{_{D}})^{3}} \right
    ] \; ,
\end{equation}
\begin{equation}
 \frac{d\rho_{_{L}}}{dr}= \frac{{M}_{_{L}}}{R_{e_{L}}^{4}} \frac{b^{2n}}{2 \pi n \Gamma(2n)}\frac{d\widetilde{\rho}}{dx}|_{x=r/R_{e_{L}}} \; ,
\end{equation}
\begin{equation}
 \frac{d\rho_{_{D}}}{dr}= \frac{{M}_{_{L}}}{R_{e_{L}}^{4}} \frac{\mu}{x_{_{D}}^{4}}\frac{b^{2m}}{2 \pi m \Gamma(2m)}\frac{d\widetilde{\rho}}{dx}|_{x=x/x_{_{D}}} \; ,
\end{equation}
\begin{equation}
 \frac{d^{2}\rho_{_{L}}}{dr^{2}}= \frac{{M}_{_{L}}}{R_{e_{L}}^{5}} \frac{b^{2n}}{2 \pi n \Gamma(2n)}\frac{d^{2}\widetilde{\rho}}{dx^{2}}|_{x=r/R_{e_{L}}} 
\end{equation}
and
\begin{equation}
 \frac{d^{2}\rho_{_{D}}}{dr^{2}}= \frac{{M}_{_{L}}}{R_{e_{L}}^{5}} \frac{\mu}{x_{_{D}}^{5}}\frac{b^{2m}}{2 \pi m \Gamma(2m)}\frac{d^{2}\widetilde{\rho}}{dx^{2}}|_{x=x/x_{_{D}}} \; ,
\end{equation}
These equations show that the $f(\mathcal{E})$ is completely defined by the first and second derivatives of the density profile, the gravitational potential and the mass profiles of the two Sersic components. In order to calculate $f(\mathcal{E})$, we derive numerically the functions $\widetilde{\rho}$, $\frac{d \widetilde{\rho}}{d r}$, $\frac{d^{2} \widetilde{\rho}}{d r^{2}}$, $\widetilde{\phi}$, and $\widetilde{M}$, and then, using Eq.~\ref{eq_eddington_finale}, we evaluate Eq.~\ref{eq_distribution_function}.

To prove the stability of the double Sersic models, one has to prove the condition
$\frac{df}{d \varepsilon} \geq 0$  (see Sec.~\ref{consistency}). From the Eddington formula, a necessary condition is $\frac{d^{2} \rho}{d \Psi^{2}} \ge 0$.
From Eq.~\ref{d2rd2p}, this condition can be written as
\begin{eqnarray}
 \frac{d^{2} \rho}{d r^{2}}
    \left ( \frac{d \Psi}{d r} \right )^{-2} - \frac{d \rho}{d r}
    \left( \frac{d \Psi}{d r} \right
    )^{-3} \cdot \frac{d^{2} \Psi}{dr^{2}}   &=&
    \left ( \frac{d \Psi}{d r} \right
    )^{-3} \left[ \frac{d^{2} \rho}{d r^{2}} \left ( \frac{d \Psi}{d r} \right
    ) - \left(\frac{d \rho}{d r}\right) \frac{d^{2} \Psi}{dr^{2}} \right] \geq 0
    \; .
\end{eqnarray}
Since $\frac{d \Psi}{dr}$ is negative (i.e. the gravitational potential is a monotonically increasing function of $r$), the previous condition is equivalent to :
\begin{equation}
    g(r; n, \mu, x_{_{D}}) = - \left[ \frac{d^{2} \rho}{d r^{2}} \left ( \frac{d \Psi}{d r} \right
    ) - \left(\frac{d \rho}{d r}\right) \frac{d^{2} \Psi}{dr^{2}}
    \right] \geq 0 \; ,
\end{equation}
as stated in Sec.~\ref{DF}.

\end{document}